\documentclass[usenatbib,referee]{mn2e}
\usepackage{epsf}
\usepackage{graphicx}

\def\eps@scaling{1.0}%
\newcommand\epsscale[1]{\gdef\eps@scaling{#1}}%
\newcommand\plotone[1]{%
 \centering
 \leavevmode
 \includegraphics[width={\eps@scaling\columnwidth}]{#1}%
}%

\def\beq{\begin{equation}}
\def\eeq{\end{equation}}
\def\bey{\begin{eqnarray}}
\def\eey{\end{eqnarray}}

\def\mpc{\, h^{-1}{\rm {Mpc}}}

\def\kpc{\, h^{-1}{\rm {kpc}}}

\def\msun{\, h^{-1}{\rm M_\odot}}

\def\gs{\mathrel{\raise1.16pt\hbox{$>$}\kern-7.0pt
\lower3.06pt\hbox{{$\scriptstyle \sim$}}}}
\def\ls{\mathrel{\raise1.16pt\hbox{$<$}\kern-7.0pt
\lower3.06pt\hbox{{$\scriptstyle \sim$}}}}
\def\gtsima{$\; \buildrel > \over \sim \;$}
\def\ltsima{$\; \buildrel < \over \sim \;$}
\def\prosima{$\; \buildrel \propto \over \sim \;$}
\def\gsim{\lower.5ex\hbox{\gtsima}}
\def\lsim{\lower.5ex\hbox{\ltsima}}
\def\simgt{\lower.5ex\hbox{\gtsima}}
\def\simlt{\lower.5ex\hbox{\ltsima}}
\def\simpr{\lower.5ex\hbox{\prosima}}
\def\la{\lsim}
\def\ga{\gsim}

\title
[Ejected Halos] {The distribution of ejected subhalos and its
implication for halo assembly bias}
\author[Huiyuan Wang et al.]
   {\parbox[t]{\textwidth}{
       Huiyuan Wang$^{1,4}$\thanks{E-mail: whywang@mail.ustc.edu.cn},
       H. J. Mo$^{2}$,
       Y.P. Jing$^{3,4}$
}\\
            $^1$Center for Astrophysics, University of Science and
                Technology of China, Hefei, Anhui 230026, China\\
            $^2$Department of Astronomy, University of Massachusetts,
            Amherst MA 01003-9305, USA\\
            $^3$Shanghai Astronomical Observatory; the Partner Group of MPA,
            Nandan Road 80, Shanghai 200030, China \\
            $^4$Joint Institute for Galaxy and Cosmology (JOINGC) of SHAO
            and USTC
}
\date{
Accepted ........ Received .......; in original form ......}

\pubyear{2008}

\begin{document}
\maketitle \label{firstpage}

\begin{abstract}
Using a high-resolution cosmological $N$-body simulation, we
identify the ejected population of subhalos, which are halos at
redshift $z=0$ but were once contained in more massive `host'
halos at high redshifts. The fraction of the ejected subhalos in
the total halo population of the same mass ranges from 9\% to 4\%
for halo masses from $\sim 10^{11}$ to $\sim 10^{12}\msun$. Most
of the ejected subhalos are distributed within 4 times the virial
radius of their hosts. These ejected subhalos have distinct
velocity distribution around their hosts in comparison to normal
halos. The number of subhalos ejected from a host of given mass
increases with the assembly redshift of the host. Ejected subhalos
in general reside in high-density regions, and have a much higher
bias parameter than normal halos of the same mass. They also have
earlier assembly times, so that they contribute to the assembly
bias of dark matter halos seen in cosmological simulations.
However, the assembly bias is {\it not} dominated by the ejected
population, indicating that large-scale environmental effects
on normal halos are the main source for the assembly bias.
\end{abstract}


\section{Introduction}
In the standard cold  dark matter (CDM) paradigm of structure
formation, galaxies are supposed to form and evolve in dark matter
halos. The study of the clustering properties of dark matter halos
and their relation to galaxy clustering can thus help us to
understand the connection between halos and galaxies, and hence to
understand how galaxies form and evolve in dark matter halos. It
is now well known that the correlation strength of dark matter
halos depends strongly on halo mass (e.g., Mo \& White 1996;
Mo et al. 1997; Jing
1998; Sheth \& Tormen 1998; Sheth, Mo \& Tormen 2001; Seljak \&
Warren 2004), and this dependence, which is referred to as the
halo bias,  has been widely used to understand the clustering of
galaxies via the halo occupation model (e.g., Jing, Mo \&
B\"{o}rner 1998; Peacock \& Smith 2000), and the conditional
luminosity function model (e.g., Yang, Mo \& van den Bosch 2003).

More recently, a number of independent investigations have shown
that the halo bias depends not only on the mass but also assembly
time of dark matter halos, in the sense that halos of a given
mass, particularly low-mass ones, are more strongly correlated if
they assembled half of their masses earlier (e.g. Gao et al. 2005;
Harker et al. 2006; Zhu et al. 2006; Wechsler et al. 2006; Jing,
Suto \& Mo 2007; Wetzel et al. 2007; Bett et al. 2007; Gao et al.
2007; Li et al. 2008). The origin of this assembly-time dependence
of the halo bias, referred to in the literature as the halo
assembly bias, is important to understand, because if it is due to
a process that can also affect galaxy formation and evolution in a
halo then it would affect our interpretation of galaxy clustering
in terms of halo clustering. There have been theoretical
investigations about the origin of the halo assembly bias (e.g.
Wang et al. 2007; Sandvik et al. 2007; Desjacques 2007; Keselman
\& Nusser 2007; Dalal et al. 2008; Hahn et al. 2008). Wang et al.
(2007) find that old, small halos tend to live near massive halos,
and suggest that the tidal truncation of accretion may be
responsible for the assembly bias. This is consistent with the
result of Maulbetsch et al. (2007), who find that most of the
small halos in high-density regions have ceased accretion. Along
this line, Desjacques (2007) develops an ellipsoidal collapse
model which takes into account the large-scale tidal field, while
Keselman \& Nusser (2007) perform simulations using the Zel'dovich
(PZ) approximation to take into account the large-scale tidal
effects. Both investigations find significant dependence of halo
bias on halo assembly history, indicating that large-scale tidal
effects may play an important role in producing the assembly bias.
More recently, Ludlow et al. (2008, hereafter L08) study in detail
5 simulations of dark matter halos and find that a significant
fraction of small halos are physically associated with nearby
massive halos. These small halos have been once inside their host
halos but were ejected due to interactions with other subhalos
(see also Lin et al. 2003; Wang et al. 2005; Gill, Knebe, \&
Gibson 2005). L08 suggest that these ejected subhalos may be
responsible for the enhanced clustering of old small halos.
However, because of the small volume of their simulations, they
were not able to quantify whether the ejected population alone can
account for the assembly bias seen in cosmological simulations.

In this paper we use a high-resolution $N$-body simulation in a
cosmological box to study the distribution of ejected subhalos in
space and to quantify the contribution of this population of halos
to the assembly bias. The outline of the paper is as follows. In
Section \ref{sec_sim} we describe briefly the simulation used and
how ejected subhalos are identified. In Section \ref{sec_rel} we
study the distribution of the ejected subhalos in phase space, and
how the distribution depends on the properties of their hosts. In
Section \ref{sec_ori} we examine the contribution of the ejected
subhalos to the assembly bias obtained in our simulation. Finally,
in Section \ref{sec_dis}, we discuss and summarize our results.

\section{Simulation and Dark Matter halos}\label{sec_sim}

\subsection{Numerical simulation and halo identification}

The simulation used in this paper is obtained with the ${\rm
P^3M}$ code described in Jing \& Suto (2002). It assumes  a
spatially-flat $\Lambda$CDM model, with density parameters
$\Omega_{\rm m}=0.3$ and $\Omega_\Lambda=0.7$. The CDM
power spectrum is assumed to be that given by
Bardeen et al. (1986) with a shape parameter
$\Gamma=\Omega_{\rm m}h=0.2$ and an amplitude specified by
$\sigma_8=0.9$. The CDM density field is traced with $512^3$
particles, each having a mass $M_p\sim6.2 \times 10^{8}$$\msun$,
in a cubic box of 100 $\mpc$. The softening length is  $\sim 10
\kpc$ (S2 type). The simulation, started at redshift 72, is
evolved with 5000 time steps to the present day ($z = 0$) and has
60 outputs from $z=15$, equally spaced in $\log (1+z)$. Dark
matter halos were identified with a friends-of-friends algorithm
with a link length that is $0.2$ times the  mean particle
separation.

\subsection{Ejected subhalos}

Our analysis focuses on the ejected subhalos which are identified
as FOF halos at redshift $z=0$. In order to determine whether a
halo is an ejected subhalo or a normal halo, a detailed merging
tree for each FOF halo is required so that we can trace a FOF halo
back in time to see whether it has ever been within another FOF
halo. We consider a halo at any given redshift $z>0$ to be a
progenitor of a descendant halo in the next output, if more than
half of its particles are found in the descendant. A given halo in
general can have one or more progenitors but only one descendant.
We can therefore use the uniqueness of the descendant to build up
the merging tree for each halo. Each FOF halo at the present time
has one and only one merging tree to describe its assembly
history. There is a small fraction of halos at $z>0$ that have
dispersed and have no descendant at $z=0$. These halos are
excluded in our analysis. The merging tree of a halo can be used
to verify whether an isolated FOF halo (called halo `A') at $z=0$
was accreted into a massive halo earlier. To do this, we search in
the next snapshot (in reverse order of time) a `host' halo that
contains at least half of the particles in halo `A', but does not
belong to the merging tree of halo `A'.  If no such a `host' halo
is found in this snapshot, we take the most massive progenitor of
halo `A' in this snapshot and repeat the same procedure for this
progenitor as we have carried out for halo `A'. This procedure is
continued until a `host' halo is found or the tree ends. If such a
`host' halo is found, halo `A' is then identified as an ejected
subhalo, which is said to be ejected at the time, $t_{\rm e}$,
when the `host' halo is found. Note that the ejection time,
$t_{\rm e}$, is defined to be the time of the snapshot at which
the `host' and `ejected' halos were just separated from the same
FOF group. We also define an accretion time, $t_{\rm a}$, as the
time when half of the particles in the most massive progenitor of
halo `A' at ejection time is first contained in the most massive
progenitor of its `host'. If no `host' halo is found before the
merging tree ends, halo `A' is said to be a normal halo. Applying
this method to all halos at $z=0$, we construct a catalogue of
ejected subhalos, and the total number of ejected subhalos is
listed in Table \ref{tab_nh} in three mass ranges. We denote the
masses of an ejected subhalo and of the corresponding `host' halo
at any time $t$ as $M_{\rm s}(t)$ and $M_{\rm host}(t)$,
respectively, and we use $M_{\rm h}$ to denote the mass of a
normal halo at redshift $z=0$. Thus, $M_{\rm host}(t_{\rm e})$ is
the mass of the `host' halo at the time of ejection, and $M_{\rm
host}(t_0)$ is the mass of the descendant of the `host' at the
present time, $t_0$.

\begin{figure}
\epsscale{1}\plotone{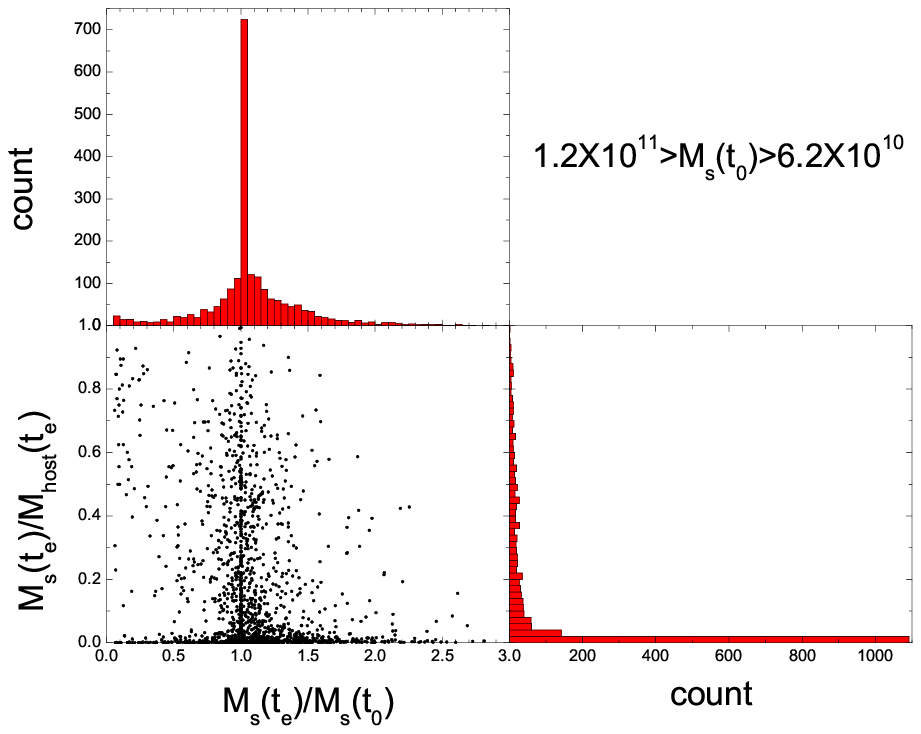} \caption{The lower-right panel
shows $M_{\rm s}(t_{\rm e})/M_{\rm host}(t_{\rm e})$ as a function
of $M_{\rm s}(t_{\rm e})/M_{\rm s}(t_0)$ for ejected subhalos with
masses between $6.2\times10^{10}$ and $1.2\times10^{11}\msun$. The
upper and lower-right panels are the histograms for $M_{\rm
s}(t_{\rm e})/M_{\rm s}(t_0)$ and $M_{\rm s}(t_{\rm e})/M_{\rm
host}(t_{\rm e})$, respectively.} \label{fig_r12}
\end{figure}

In Fig. \ref{fig_r12} we show $M_{\rm s}(t_{\rm e})/M_{\rm
host}(t_{\rm e})$ versus $M_{\rm s}(t_{\rm e})/M_{\rm s}(t_0)$ for
halos with $6.2\times10^{10}<M_{\rm
s}(t_0)<1.2\times10^{11}\msun$. The corresponding histograms for
$M_{\rm s}(t_{\rm e})/M_{\rm s}(t_0)$ and $M_{\rm s}(t_{\rm
e})/M_{\rm host}(t_{\rm e})$ are also shown in the figure. The
results for other mass ranges are similar and are not shown. As
one can see, the majority of the ejected subhalos were ejected by
`host' halos that are much more massive than the ejected subhalos
themselves. Only small fraction of halos are ejected by host halos
with comparable masses: about 11\% (34\%) of ejected subhalos have
a mass larger than 0.5  (0.1) times of their host halo mass.
Furthermore, most of the ejected halos have masses that are
similar to those at the time of ejection, indicating that mass
loss or accretion after ejection is not severe. However, the
distribution has two extended tails, and the 10th and 90th
percentile values are 0.73 and 1.48, respectively. In our
analysis, we remove all systems with $M_{\rm s}(t_{\rm e})/M_{\rm
s}(t_0)< 0.5$, i.e. the halos that have accreted more than their
initial masses after ejection. Note that the removed halos is a
very small fraction, about 5\%, of the total sample (Table
\ref{tab_nh}). Including them does not have any significant impact
on the results to be presented below. After the removal, the final
numbers of the ejected subhalos in different halo mass ranges are
also given in Table \ref{tab_nh}. For comparison, we also list the
corresponding numbers for all (ejected plus normal) halos in these
mass ranges. About 9\% - 4\% of all the halos in the mass ranges
considered are ejected subhalos, and the fraction decreases with
increasing $M_{\rm s} (t_0)$.

\begin{table}
\begin{center}
\caption{The number of ejected subhalos in various samples}
\label{tab_nh}
\begin{tabular}{lccc}
  \hline\hline
  % after \\: \hline or \cline{col1-col2} \cline{col3-col4} ...
  $M_{\rm s}/(\msun)$ & $[6.2\times10^{10}, 1.2\times10^{11}]$
                      & $[1.2\times10^{11}, 3.7\times10^{11}]$
                      & $[3.7\times10^{11}, 10^{12}]$ \\
  \hline
  Ejected (total)       & 2115  & 1076  & 231\\
  Ejected (final)       & 2009  & 1036  & 220\\
  Ejected$+$Normal      & 22697 & 16490 & 5850\\
  \hline
\end{tabular}
\end{center}
\end{table}

\begin{figure}
\epsscale{1}\plotone{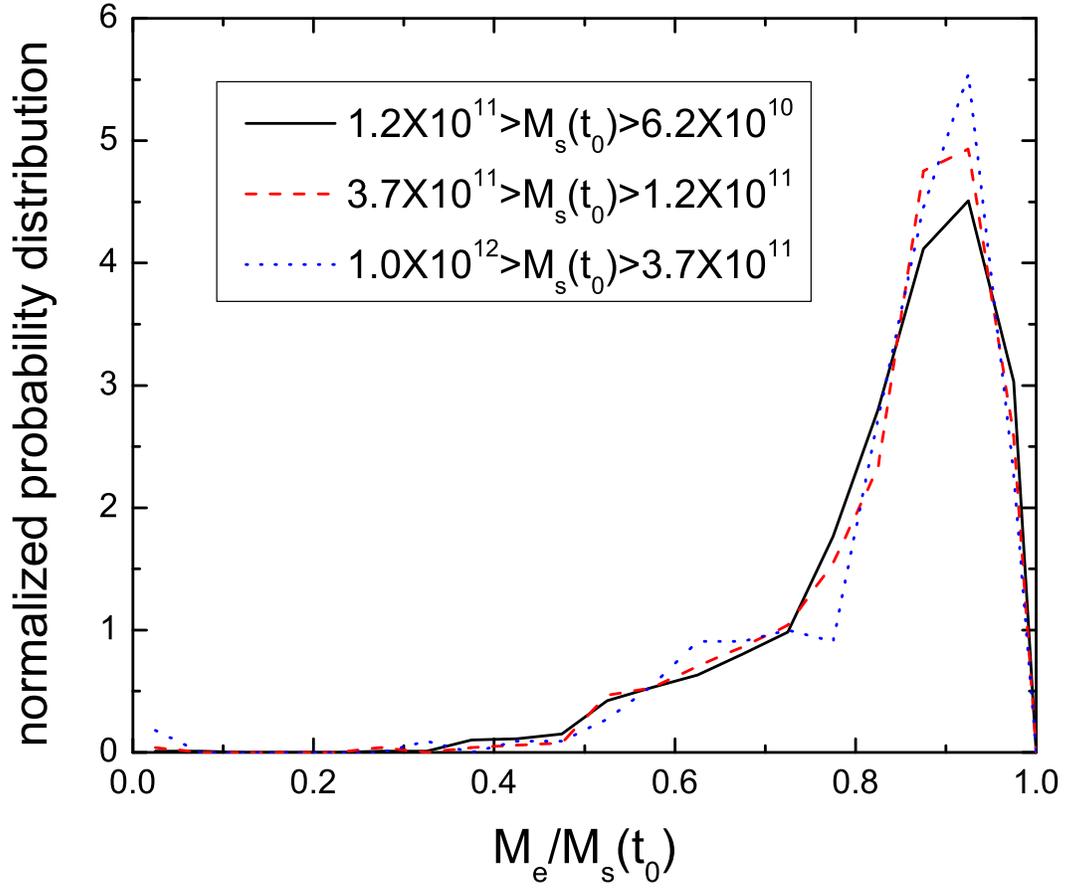} \caption{The probability
distribution of $M_{\rm e}/M_{\rm s}(t_0)$ for ejected
subhalos in three mass bins, as indicated. Masses
are in units of $h^{-1}{\rm M}_\odot$. See text for the
definition of $M_{\rm e}$.}
\label{fig_fm}
\end{figure}

 It is possible that an ejected subhalo after ejection can exchange
mass with nearby halos or with the background density field, so
that some of the particles contained in the ejected subhalo at the
present time may not be contained in its `host' halo. In order to
quantify this, we consider an ejected mass, $M_{\rm e}$, which is
defined as the mass of particles which are contained both in the
ejected subhalo at $z=0$ and in its `host' halo at the ejection
redshift. The ratio between $M_{\rm e}$ and $M_{\rm s}(t_0)$ can
then be used as a measure of the fraction of retained mass of an
ejected subhalo. In Fig. \ref{fig_fm} we show the probability
distribution of $M_{\rm e}/M_{\rm s}(t_0)$. As one can see, most
of the ejected subhalos at $z=0$ can retain more than
$70\%$ of their original masses at ejection. This also shows that our
method for identifying ejected subhalos is valid, in the sense
that most of their masses were indeed once contained in their
hosts.

\subsection{Halo assembly times}
\label{sec_ass}

We define the assembly redshift, $z_{\rm f}$, of a halo at
redshift $z=0$ as the redshift when its most massive progenitor
first reaches half of the final mass of the halo. If necessary,
interpolation between two adjacent outputs is used. Here we do not
use the merging tree described above to search for the progenitor.
Instead we use the method described in Wang et al. (2007; see also
Hahn et al. 2008). Very briefly, a halo in any given output at
$z>0$ is considered to be a progenitor of the halo at $z=0$ if
more than half of its particles are found in the final halo. In
most cases, the merging trees constructed with this method is very
similar to that constructed with the method described above. The
advantage of the method adopted here is that it can trace an
ejected subhalo backwards to the time even before it is accreted
by its host halo. Since ejected halos may be strongly stripped by
their host halos and lose part of their masses, the assembly
redshifts of the ejected subhalos are generally higher than
than their ejection redshifts and can be identified with this
method.

\section{The relationship between ejected subhalos and their hosts}
\label{sec_rel}

\begin{figure}
\epsscale{1}\plotone{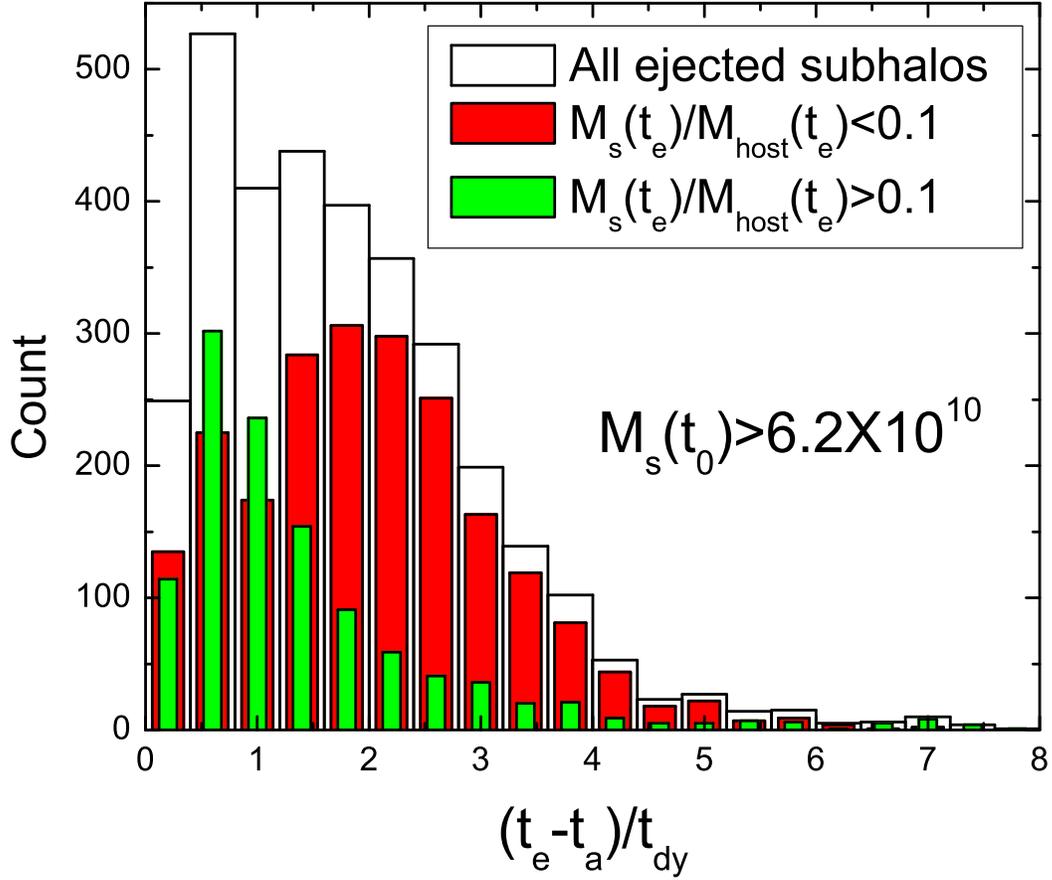}\caption{The histograms for
$(t_{\rm a}-t_{\rm e})/t_{\rm dy}$ of ejected subhalos. The white
histograms show the result for ejected subhalos with masses
$M_{\rm s}(t_0)> 6.2\times 10^{10}h^{-1}{\rm M}_\odot$. And red
and green hisograms show the results for two subsample indicated
in the panel.} \label{fig_nt}
\end{figure}

In Fig. \ref{fig_nt}, we show the histograms of the scaled time
period an ejected subhalo stayed in its host: $(t_{\rm e}-t_{\rm
a})/t_{\rm dy}$. Here $t_{\rm dy}$ is the dynamical time at
$r_{200}$ of each host halo at the ejection time, with $r_{200}$
the virial radius within which the mean overdensity of the halois
200 times the critical density. The dynamical time is defined as:
\begin{equation}
t_{\rm dy}=\frac{r_{200}(t_{\rm e})}{v_{200}(t_{\rm e})}
=\frac{1}{10H(t_{\rm e})}\,,
\end{equation}
where $v_{200}$ is the circular velocity at $r_{200}$ and
$H(t_{\rm e})$ is the Hubble constant, at the ejection time. As
one can see, the time period an ejected subhalo stays within its
host ranges from $<0.5$ to $>4$ times the dynamical time. To
examine this time period in more detail, we split the ejected
subhalos sample into two subsamples according to the $M_{\rm
s}(t_{\rm e})/M_{\rm host}(t_{\rm e})$ ratio. The red (dark) and
green (light) histograms in Fig. \ref{fig_nt} represent the
results with $M_{\rm s}(t_{\rm e})/M_{\rm host}(t_{\rm e})<0.1$
and $>0.1$, respectively. Evidently, the distribution for the case
with $M_{\rm s}(t_{\rm e})/M_{\rm host} (t_{\rm e})<0.1$ is peaked
around 2, and only 24\% of halos have $(t_{\rm e}-t_{\rm
a})/t_{\rm dy}<1$. On the other hand, halos with large $M_{\rm s}
(t_{\rm e})/M_{\rm host} (t_{\rm e})$ ratio show quite a different
distribution. It peaks at $\sim$ 0.6, and the fraction of halos
with $(t_{\rm e}-t_{\rm a})/t_{\rm dy}< 1$ is more than 55\%. This
indicates that many of the ejected subhalos with a large $M_{\rm
s} (t_{\rm e})/M_{\rm host} (t_{\rm e})$ may be flybys which
happen to be close to another halo and be linked to it by the FOF
group finder. They may, therefore, represent a population that is
different from the population with a small $M_{\rm s} (t_{\rm
e})/M_{\rm host} (t_{\rm e})$, which most likely have run through
their hosts. In what follows, we will treat these two populations
separately. It should be pointed out, however, that there is an
excess at 0.6 for halos with $M_{\rm s}(t_{\rm e})/M_{\rm
host}(t_{\rm e})<0.1$, and there is a long tail for the subsample
with $M_{\rm s}(t_{\rm e})/M_{\rm host}(t_{\rm e})>0.1$ (see Fig.
\ref{fig_nt}). Thus, the $M_{\rm s} (t_{\rm e})/M_{\rm host}$
ratio alone cannot distinguish the two populations unambiguously.

\begin{figure}
\epsscale{1}\plotone{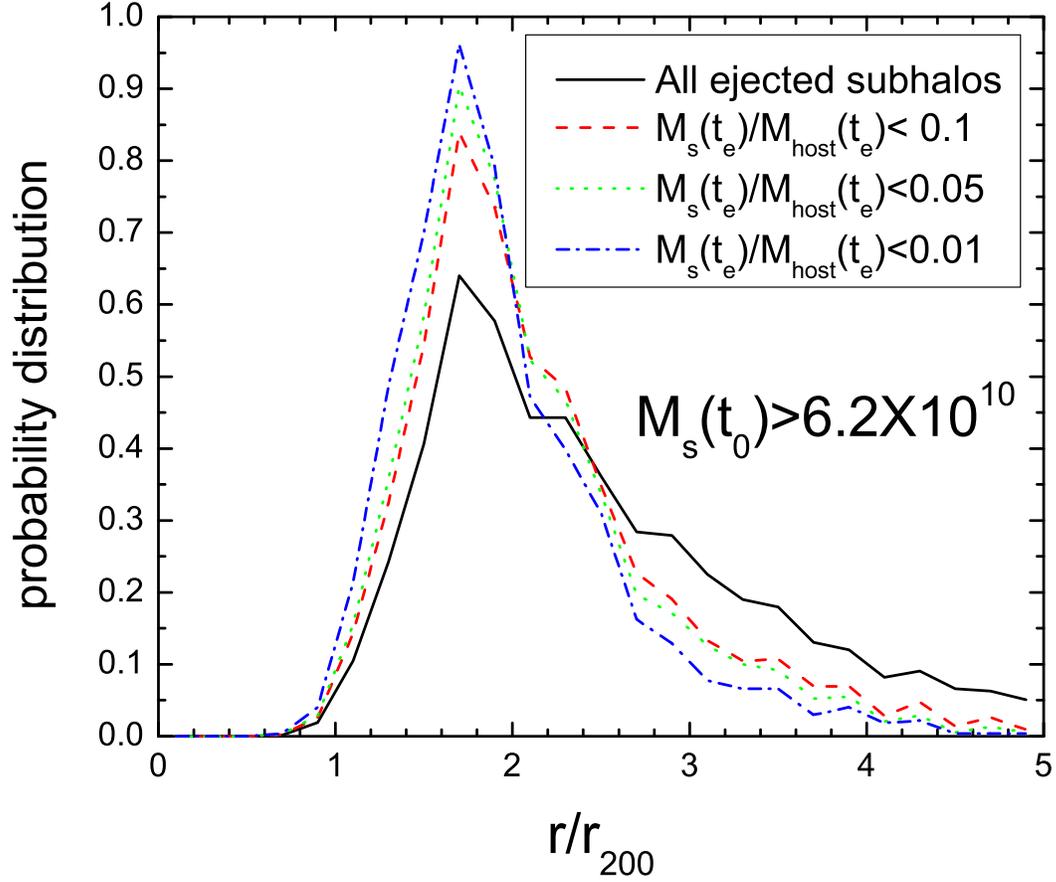}\caption{The probability
distribution of ejected subhalos [with masses $M_{\rm s}(t_0)>
6.2\times 10^{10}h^{-1}{\rm M}_\odot$] in their distances to the
host halos at $z=0$. The solid line shows the result for all
ejected subhalos. The other lines show the results for the ejected
halos with $M_{\rm s}(t_{\rm e})/M_{\rm host}(t_{\rm
e})<0.1,0.05,0.01$ as indicated in the panel. The distance is
scaled by $r_{200}$ of each host.} \label{fig_nrv}
\end{figure}

We show the probability distribution of ejected subhalos in their
distances to the host halos at $z=0$ in Fig. \ref{fig_nrv}. Here,
the distance is scaled by $r_{200}$ of each host halo at redshift
$z=0$. We use the scaled distance $r/r_{200}$ instead of $r$,
because $r_{200}$ is the only important length scale related to
the dynamics of a virialized halo. As one can see,  the
distribution peaks at $r/r_{200}\sim 1.6$. This is different from
what is shown in L08, because their result includes also subhalos
within host halos. Most of the ejected subhalos are distributed
within $4r_{200}$, in good agreement with the finding of L08, but
the distribution has a long tail at large distance. Our detailed
examination shows that the long tail is dominated by ejected
subhalos that have masses not much smaller than those of their
`hosts' at ejection. If we \emph{exclude} the ejected subhalos
with $M_{\rm s} (t_{\rm e})/M_{\rm host} (t_{\rm e})>\mu$, then
the extended tail disappears as long as $\mu$ is chosen to be
smaller than $0.1$. The distribution obtained for $\mu=0.1$,
$0.05$ and $0.01$ are shown in Fig.\ref{fig_nrv} for comparison.
This result indicates that the two populations of the ejected
subhalos discussed above have different distribution. Ludlow et
al.(2008) found that subhalos with small mass at accretion time is
less centrally concentrated than their massive counterparts. We do
not find any significant evidence for such dependence. The
difference in the result may be due to the difference in the halo
definition and, particularly, in the contamination of the flybys
population found in our simulation. This flyby population may be
not important in L08 because of their way of selecting halos (see
section 2.2 in their paper).

\begin{figure}
\epsscale{0.5}\plotone{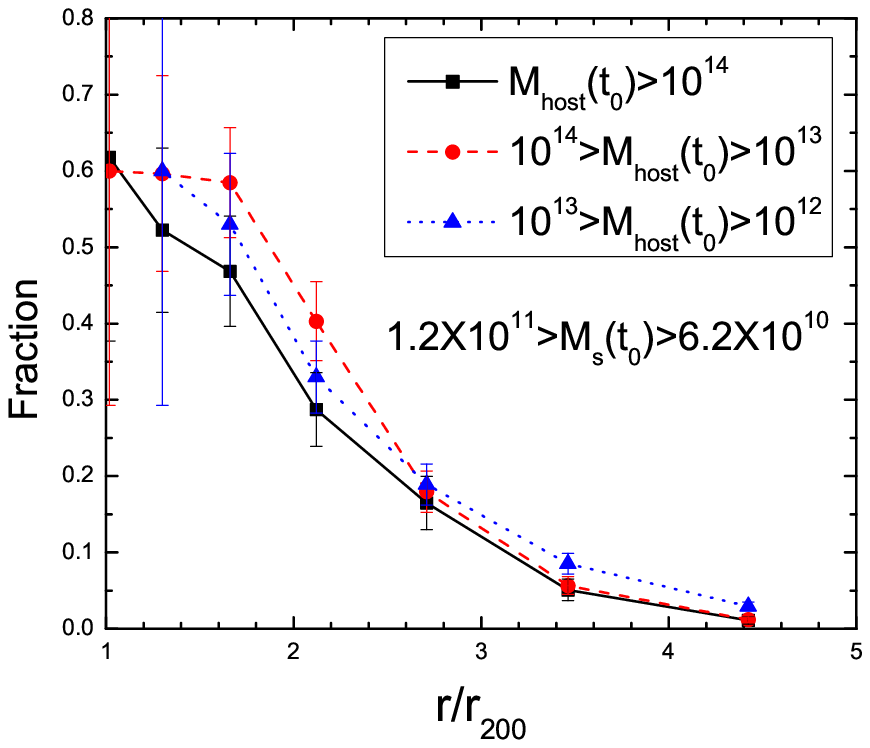}\plotone{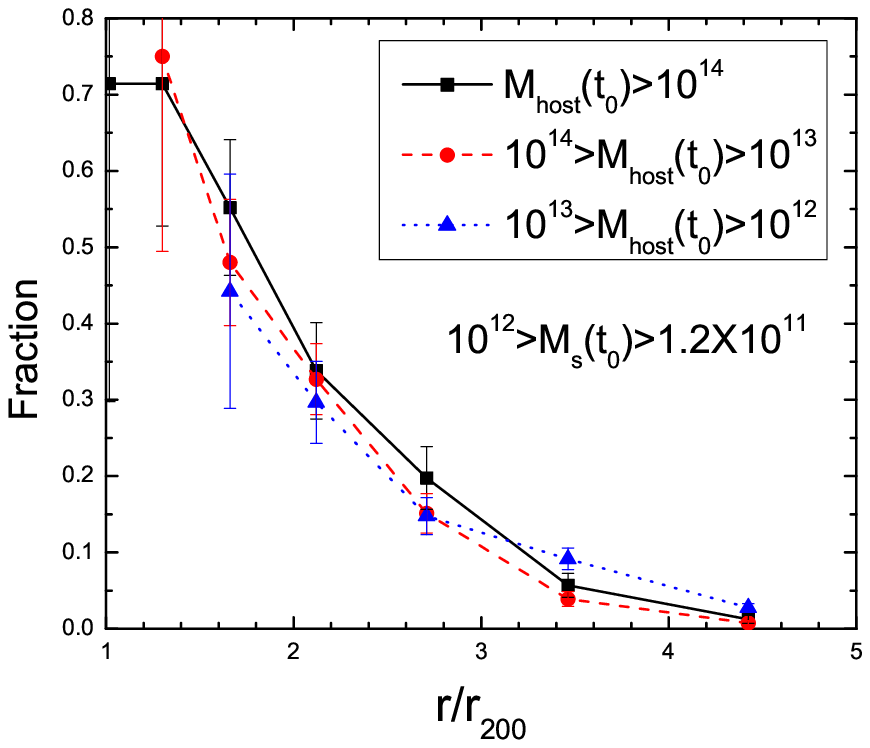}
\caption{The ratio between the number of ejected subhalos and that
of the total population (ejected plus normal) as a function of the
distance to the hosts in various mass ranges.
All masses are in $h^{-1}{\rm M}_\odot$.
The distance is scaled by $r_{200}$ of each host.} \label{fig_frv}
\end{figure}

Ejected subhalos in general are distributed around massive halos,
but there are also normal halos which are distributed near massive
halos but have never been inside a massive halo. In Fig.
\ref{fig_frv}, we show the average fraction of ejected subhalos
around their host halos. This is the ratio between the number of
ejected subhalos and that of the total population (ejected plus
normal) as a function of the scaled distance to the host halos.
The left panel shows the average fractions of ejected subhalos
with $6.2\times10^{10}<M_{\rm s}(t_0)<1.2\times10^{11}\msun$
around host halos in three mass ranges. In the right panel, we
show the results for ejected subhalos with
$1.2\times10^{11}<M_{\rm s}(t_0)<10^{12}\msun$.
In the mass ranges probed here, the ejected fraction as a function
of the scaled distance is insensitive to the masses of both
the ejected subhalos and the host halos. The fraction is between 30\% and
75\% in the range $r_{200}<r<2r_{200}$ and between 10\% and 40\%
in the range $2r_{200}<r<3r_{200}$. These results are in agreement
with those of L08, who found fractions of about 65\% and 33\%
in the similar distance ranges (see also Gill et al. 2005; Wang
et al. 2005), although the mass ranges for both the host halo and
the ejected subhalos considered by them are quite different from
what we are considering here. Note that, at large scaled distances, the
fraction of ejected subhalos around host halos with the lowest
mass is larger than that around massive host halos. This is again
due to the population in which the ejected subhalos have masses
comparable to their hosts at the time of ejection.

\begin{figure}
\epsscale{0.5}\plotone{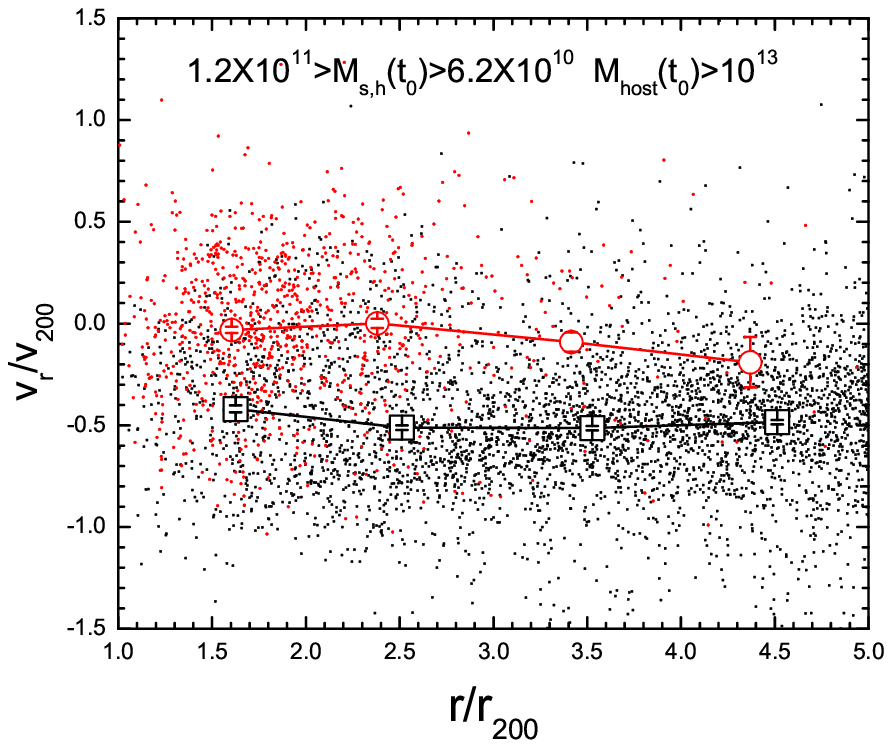}\plotone{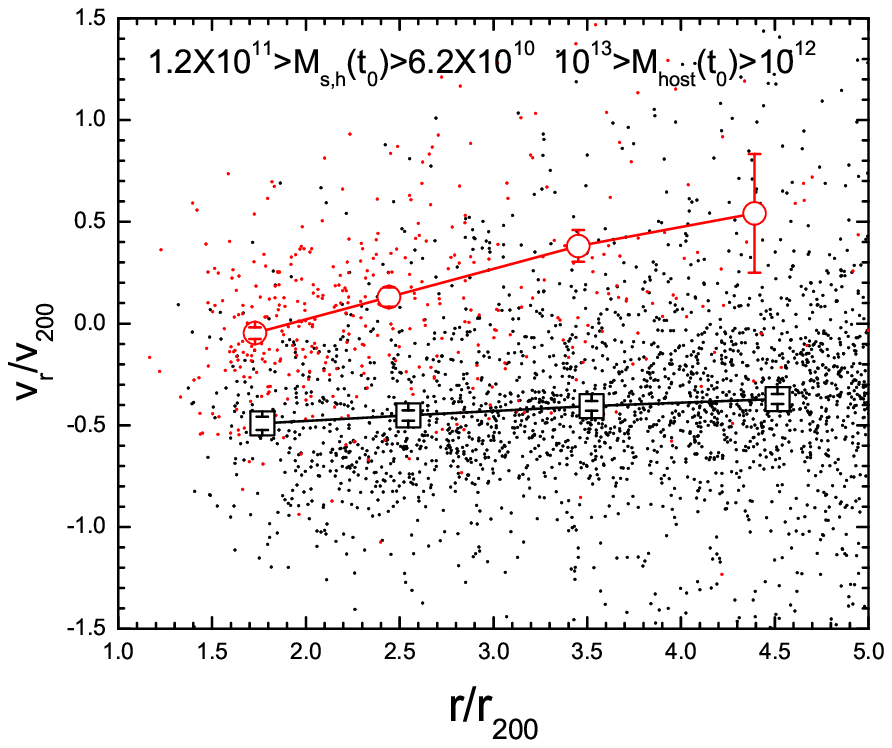}
\caption{Radial velocity versus distance to the host halo for both
ejected subhalos (red) and normal halos (black) in the mass ranges
as indicated in the panels (all masses are in $h^{-1}{\rm M}_\odot$).
The left panel shows halos around
host halos with masses above $10^{13}\msun$ and the right panel
shows those around hosts with masses between $10^{13}\msun$ and
$10^{12}\msun$. The symbols with error bars show the average
radial velocities in different radius bins for both ejected
subhalos (red circles) and normal halos (black squares). Note that
the radial velocity shown here is the peculiar velocity relative
to the host.} \label{fig_vrr}
\end{figure}

In Fig. \ref{fig_vrr} we show the scaled radial velocity, $v_{\rm
r}/v_{200}$, versus scaled distance for both ejected and normal
halos within $5\times r_{200}$ of the host halos of ejected halos.
Here $v_{200}$ is the circular velocity of each host halo, and
$v_{\rm r}$ is the relative \emph{peculiar} velocity along the
separation between a host and an ejected or a normal halo. We also
split the ejected subhalos and normal halos into four subsamples
according to the scaled distance and calculate the average
velocity and average distance for each subsample. The results are
shown as the big symbols in the figure. We consider two mass
ranges of host halos, $M_{\rm host}>10^{13}\msun$ and
$10^{13}>M_{\rm host}>10^{12}\msun$. Note that some normal halos
may be counted more than once since they may be within $5\times
r_{200}$ of more than one host halo. Clearly, the ejected subhalos
and normal halos show different distribution in the phase space.
The radial velocity distribution of ejected subhalos associated
with massive hosts is quite symmetric and quite independent of the
distance to the hosts, suggesting that on average there is about
equal chance for an ejected subhalo to be moving away from or
falling back towards the host. It is also consistent with the
results of L08 although the velocities shown in their paper are
the peculiar velocities plus the Hubble flow. The average radial
velocity of ejected halos associated with low-mass hosts increase
with the distance to the hosts. This is because of the
contamination of the flybys mentioned above. The behavior of
normal halos is quite different. Most of the normal halos are
preferentially moving towards the host halos, as indicated by the
systematically negative values of $v_{\rm r}$. The average value
of $v_{\rm r}/v_{200}$ is about $-0.5$, independent on the
distance and the mass of the hosts. The difference here suggests
that the ejected subhalos are a distinctive population among the
total halo population. A careful search shows that the scaled
velocity dispersion of normal halos around small host halos is
larger than that around massive host halos. This may be due to the
fact that environmental effects play a more important role around
smaller hosts.

\begin{figure}
\epsscale{0.5}\plotone{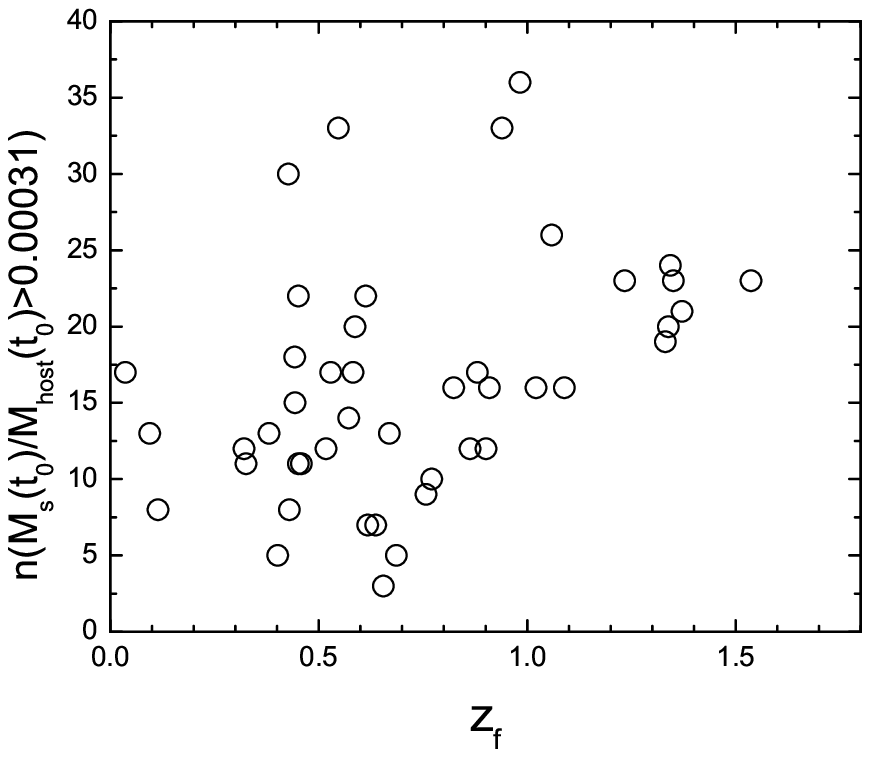}\plotone{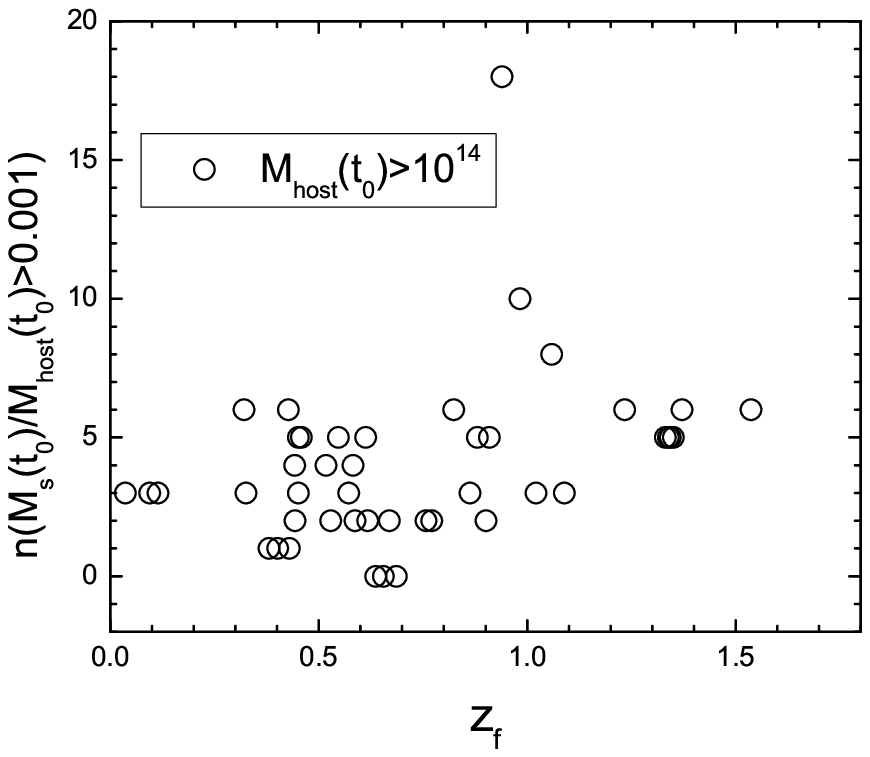}
\caption{The relationship between the number of ejected subhalos
and the assembly redshift of host halos. The left panel shows the
number of ejected subhalos with $M_{\rm s}(t_0)/M_{\rm
host}(t_0)>0.00031$ and the right panel show the results with
$M_{\rm s}(t_0)/M_{\rm host}(t_0)>0.001$. The masses of host halos
are larger than $10^{14}\msun$. } \label{fig_nm}
\end{figure}

Gao et al. (2004) find that the abundance of subhalos within host
halos decreases with the formation redshift of the hosts,
indicating that subhalos in early-formed host halos are more
likely to be disrupted. However, as pointed out by L08, the
subhalos within a host halo represent a rather incomplete census
of the substructure physically related to the host. It is thus
important to check whether the abundance of the ejected subhalos
is also related to the assembly time of their hosts. In Fig.
\ref{fig_nm}, we show the number of ejected subhalos as a function
of the assembly redshift of the host halo. The left panel is the
result for host halos with $M_{\rm host}(t_0)>10^{14}\msun$ and
$M_{\rm s}(t_0)/M_{\rm host}(t_0)>0.00031$. As one can see, there
is a clear trend that older host halos tend to eject more
subhalos, although the scatter is quite large. To compare with Gao
et al. (2004), we also show the results with $M_{\rm
s}(t_0)/M_{\rm host}(t_0)>0.001$ in the right panel of
Fig.\ref{fig_nm}. The trend is weaker, likely due to the
statistical variation caused by small numbers. This trend suggests
that the old host halos tend not only to destroy their subhalos,
but also to eject them. The decrease of subhalo abundance with
formation redshift observed by Gao et al. (2004) is therefore a
result of both subhalo ejection and destruction. For each host,
the number of destroyed subhalos, $N_{\rm d}$, reads,
\begin{equation}
N_{\rm d}=N_{\rm a}-(N_{\rm s}+N_{\rm e})\,,
\end{equation}
where $N_{\rm a}$ is the number of total accreted subhalos, and
$N_{\rm s}$ and $N_{\rm e}$ are the numbers of surviving subhalos
within the host and of ejected subhalos, respectively. Thus, in
order to quantify the importance of ejection and destruction, one
needs detailed merging trees to trace the evolution of subhalos
within their hosts.

\section{The origin of assembly bias}
\label{sec_ori}

\begin{figure}
\epsscale{1}\plotone{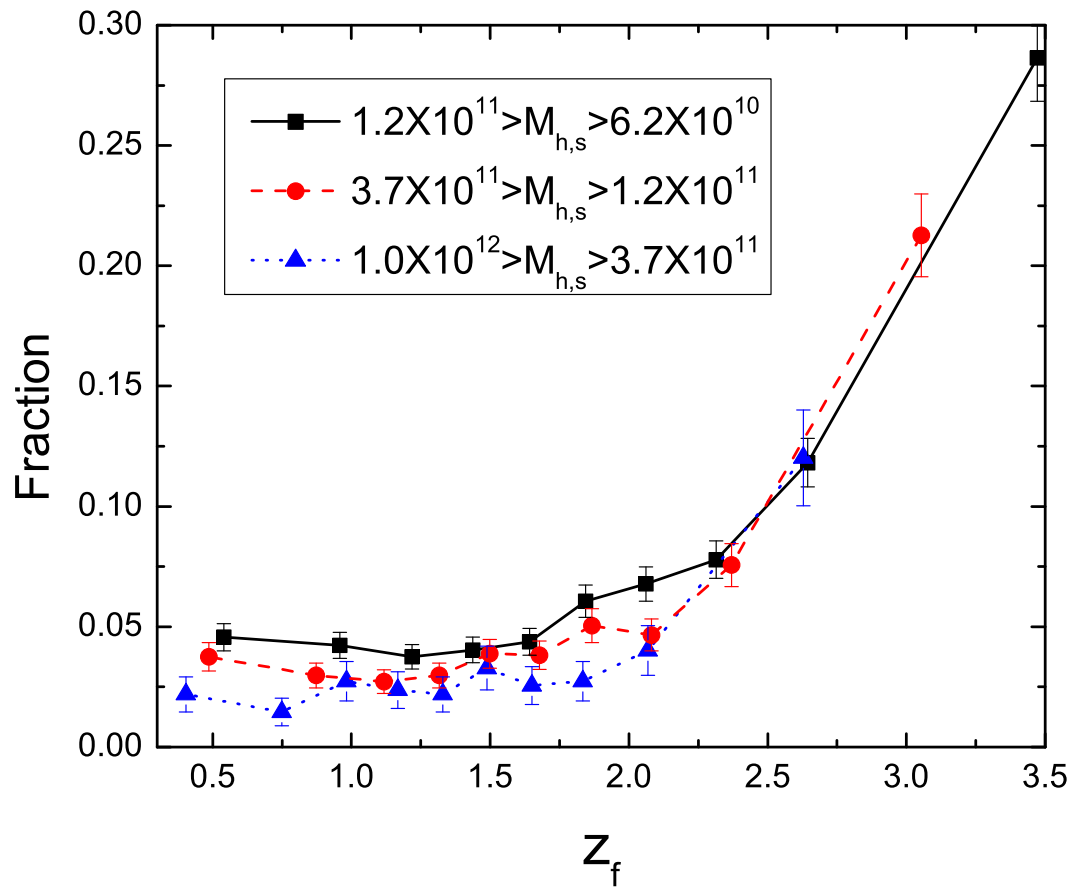} \caption{The fraction of ejected
halos as a function of assembly redshift.
All masses indicated in the panel are in $h^{-1}{\rm M}_\odot$.}
\label{fig_fz}
\end{figure}

Due to the strong interaction with their hosts before ejection,
ejected subhalos may lose part of their mass. Since it is
difficult for them to accrete more mass after ejection, the
ejected subhalos should have acquired most of their mass before
the ejection. Thus, on average these halos should have assembly
times (defined in Section \ref{sec_ass}) that are earlier than
their normal counterparts. In order to show this, we first split
the halo sample in a certain mass range into 10 equal-sized
subsamples according to assembly redshift. We then calculate the
fraction of ejected halos and the average assembly redshift for
each of these subsamples and show the fraction versus the redshift
in Fig. \ref{fig_fz}. Among the 10 percent halos with the highest
assembly redshifts, about $12\%$ to $27\%$ are ejected subhalos
depending on the subhalo mass in question. The fraction decreases
rapidly down to about $2\%$ - $5\%$ for halos with the lowest
assembly redshifts. Since the ejected subhalos are expected to be
located in high-density regions due to their associations with
massive halos, they are expected to be strongly clustered. The
presence of this population of ejected subhalos may therefore
contribute to the assembly bias seen in cosmological $N$-body
simulations (e.g. Gao et al. 2005; Jing, Suto \& Mo 2007).

\begin{figure}
\epsscale{1}\plotone{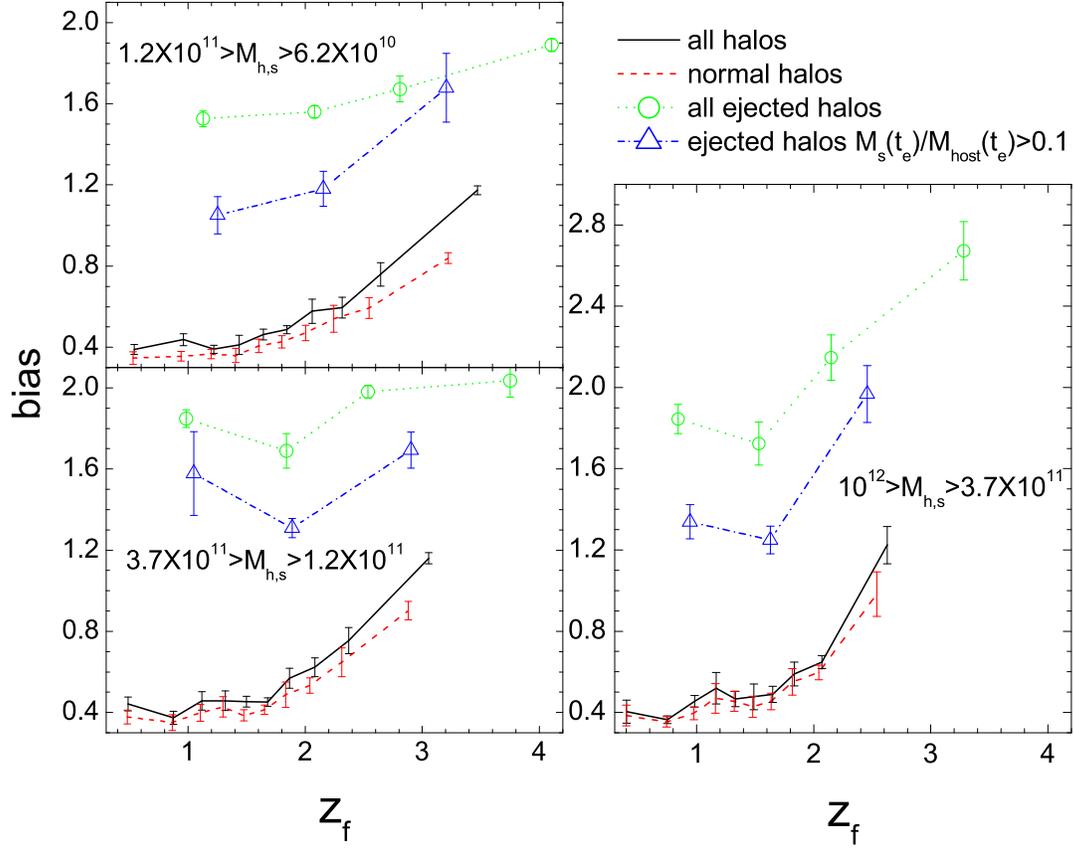} \caption{Halo bias as a function
of assembly redshift for all halos (black solid lines), normal
halos (red dash lines), and ejected subhalos (green dot lines).
All masses indicated are in units of $h^{-1}{\rm M}_\odot$.}
\label{fig_bz}
\end{figure}

In order to examine whether or not the ejected subhalos are fully
responsible for the assembly bias, we estimate the halo bias as a
function of assembly redshift separately for all halos, normal
halos and ejected (sub)halos. We first estimate the mean
overdensity of dark matter within a sphere of radius $R$ around
each dark matter halo, $\delta_h(R)$, and then measure the bias
parameter of a given set of halos using
\begin{equation}
b=\frac{\langle\delta_h(R)\rangle}{\langle\delta_m(R)\rangle}\,,
\end{equation}
where   $\langle \delta_h(R)\rangle$ is the average overdensity
around the set of halos in question, and $\langle
\delta_m(R)\rangle$ is the average overdensity within all spheres
of radius $R$ centered on dark matter particles. This method has
been demonstrated to reproduce the bias obtained using
auto-correlation function of halos (Wang et al. 2007). The results
are shown in Fig.\ref{fig_bz}. As expected, the ejected subhalos
in general have a much higher bias parameter than other halos of
similar masses. The bias parameter for ejected subhalos with $M_{\rm
s} (t_{\rm e})/M_{\rm host} (t_{\rm e})>0.1$ is lower (blue dash
dot lines), presumably because they are associated with hosts of
lower masses, as mentioned above.
However, even if all ejected subhalos are excluded,
the assembly bias is still significant (see the red dash lines in
Fig. \ref{fig_bz}). For halos with masses larger than
$10^{11.5} h^{-1}{\rm M}_\odot$, the assembly bias is almost
entirely due to normal halos instead of ejected subhalos.
Even for halos of lower masses (e.g. $\sim 10^{11} h^{-1}{\rm M}_\odot$),
including ejected only increases the bias parameter by $50\%$.
The reason is that the fraction of ejected subhalos is small
among the total halo population of the same mass.
Thus, ejected subhalos cannot explain the full
range of assembly bias seen in cosmological $N$-body simulations,
even though they are strongly clustered in space.
This result is consistent with the finding of Wang et al. (2007)
that the assembly bias is mainly due to the fact that small old
halos tend to live in the vicinity of massive halos and their
growth at low redshift is suppressed by the large-scale tidal
field. An analysis along this line is performed by Keselman \&
Nusser (2007). They used the punctuated Zel'dovich approximation,
in which the highly non-linear effects such as tidal stripping are
excluded, to run simulations of structure formation, and found
that the assembly bias is still present. All these indicate that
the large-scale environmental effects have also played an
important role in the formation of some normal halos (see also
Hahn et al. 2008).

\section{Summary and discussion}
\label{sec_dis}

In this paper, we use a high-resolution cosmological simulation to
study the distribution of the ejected subhalos, their connection
to the host halos, and their contribution to the halo assembly
bias seen in cosmological simulations. Our main results are
summarized as follows.

\begin{enumerate}

\item The fraction of the ejected subhalos in the total halo
population of the same mass ranges from 9\% to 4\% for halo masses
from $\sim 10^{11}$ to $\sim 10^{12}\msun$.

\item The time period an ejected subhalo stays in its host has
wide distribution, ranging from less than 0.5 to more than 4 times
the dynamical time of of the host. The distribution peaks at 2 and
0.6 for ejected subhalos with $M_{\rm s}(t_{\rm e})/M_{\rm
host}(t_{\rm e})<0.1$ and $>0.1$, respectively, indicating the
existence of two distinctive populations of ejected halos.

\item Most of the ejected subhalos are found to be distributed
within about 4 times the virial radius of their hosts. The fraction
of ejected subhalos is about 30\% - 75\% within the distance range
between $r_{200}$ and $2r_{200}$ to the hosts and about 10\% -
40\% in the distance range $2r_{200}<r<3r_{200}$.

\item
The radial velocity distribution of ejected subhalos
is quite symmetric, while normal halos of the
same mass generally tend to fall onto nearby massive halos.

\item The number of subhalos ejected from a host of given mass
increases with the assembly redshift of the host. Thus, subhalos
tend to be less abundant in halos that formed earlier, not only
because subhalos in such hosts are more likely to be destroyed,
but also because they are more likely to be ejected.

\item Ejected subhalos in general reside in high-density regions,
and have a much higher bias parameter than normal halos of the
same mass. They also have earlier assembly times, so that they
contribute to the assembly bias of dark matter halos seen in
cosmological simulations.

\item The assembly bias is {\it not} dominated by the ejected
population. This indicates that large-scale environmental effects
may also be important in the formation of normal halo population,
and in producing the assembly bias.

\end{enumerate}

 The results obtained here may have important implications to the
understanding of galaxy distribution in the cosmic density field.
As a subhalo pass through a massive host, tidal and/or
ram-pressure stripping by the host halo may get rid of the gas
reservoir in the ejected halo, thereby quenching star formation in
it. The situation may be similar to what happens to satellite
galaxies,  although the ejected galaxies are not observed as
satellites in massive halos. If the quenching processes are
effective, we would expect a population of faint red galaxies that
are not contained in any massive halos but are distributed around
them. It is therefore interesting to see if such a population of
galaxies does exist. In a recent investigation, Wang et al. (2008)
found that the reddest $15$ - $20\%$ among all faintest galaxies
are physically associated with massive halos. About half of this
population resides within massive halos as satellites. The other
half resides outside massive halos and are distributed within
about 3 times the virial radii of their nearest massive halos.
Very likely, this population of galaxies are hosted by the ejected
subhalos we are studying here. Clearly, it is interesting to study
further the connection between ejected subhalos and this
population of galaxies, so as to understand how environmental
effects operate on satellite galaxies in their host halos.

\section*{Acknowledgment}

We thank the anonymous referee for his/her constructive comments.
HYW would like to acknowledge the support  of the Knowledge
Innovation Program of the Chinese Academy of Sciences, Grant No.
KJCX2-YW-T05 and NSFC 10643004. HJM would like to acknowledge the
support of NSF AST-0607535, NASA AISR-126270 and NSF IIS-0611948.
YPJ is supported by NSFC (10533030, 10821302, 10878001), by the
Knowledge Innovation Program of CAS (No. KJCX2-YW-T05), and by 973
Program (N o.2007CB815402).

%%%%%%%%%%%%%%%

% Bibliography

%%%%%%%%%%%%%%%

\begin{thebibliography}{}

\bibitem[\protect\citeauthoryear{Bardeen et
al.}{1986}]{1986ApJ...304...15B} Bardeen J.~M., Bond J.~R., Kaiser
N., Szalay A.~S., 1986, ApJ, 304, 15

\bibitem[\protect\citeauthoryear{Bett et al.}{2007}]{2007MNRAS.376..215B}
Bett P., Eke V., Frenk C.~S., Jenkins A., Helly J., Navarro J.,
2007, MNRAS, 376, 215

\bibitem[\protect\citeauthoryear{Dalal et al.}{2008}]{2008arXiv0803.3453D}
Dalal N., White M., Bond J.~R., Shirokov A., 2008, arXiv,
arXiv:0803.3453

\bibitem[\protect\citeauthoryear{Desjacques}{2008}]{2008MNRAS.388..638D}
Desjacques V., 2008, MNRAS, 388, 638

\bibitem[\protect\citeauthoryear{Gao et al.}{2004}]{2004MNRAS.355..819G}
Gao L., White S.~D.~M., Jenkins A., Stoehr F., Springel V., 2004,
MNRAS, 355, 819

\bibitem[\protect\citeauthoryear{Gao, Springel, \&
White}{2005}]{2005MNRAS.363L..66G} Gao L., Springel V., White
S.~D.~M., 2005, MNRAS, 363, L66

\bibitem[\protect\citeauthoryear{Gao
\& White}{2007}]{2007MNRAS.377L...5G} Gao L., White S.~D.~M.,
2007, MNRAS, 377, L5

\bibitem[\protect\citeauthoryear{Gill, Knebe,
\& Gibson}{2005}]{2005MNRAS.356.1327G} Gill S.~P.~D., Knebe A.,
Gibson B.~K., 2005, MNRAS, 356, 1327

\bibitem[\protect\citeauthoryear{Hahn et al.}{2008}]{2008arXiv0803.4211H}
Hahn O., Porciani C., Dekel A., Carollo C.~M., 2008, arXiv,
arXiv:0803.4211

\bibitem[\protect\citeauthoryear{Harker et al.}{2006}]{2006MNRAS.367.1039H}
Harker G., Cole S., Helly J., Frenk C., Jenkins A., 2006, MNRAS,
367, 1039

\bibitem[\protect\citeauthoryear{Jing}{1998}]{1998ApJ...503L...9J} Jing
Y.~P., 1998, ApJ, 503, L9

\bibitem[\protect\citeauthoryear{Jing, Mo,
\& Boerner}{1998}]{1998ApJ...494....1J} Jing Y.~P., Mo H.~J.,
Boerner G., 1998, ApJ, 494, 1

\bibitem[\protect\citeauthoryear{Jing \& Suto}{2002}]{2002ApJ...574..538J}
Jing Y.~P., Suto Y., 2002, ApJ, 574, 538

\bibitem[\protect\citeauthoryear{Jing, Suto,
\& Mo}{2007}]{2007ApJ...657..664J} Jing Y.~P., Suto Y., Mo H.~J.,
2007, ApJ, 657, 664

\bibitem[\protect\citeauthoryear{Keselman
\& Nusser}{2007}]{2007MNRAS.382.1853K} Keselman J.~A., Nusser A.,
2007, MNRAS, 382, 1853

\bibitem[\protect\citeauthoryear{Li, Mo,
\& Gao}{2008}]{2008MNRAS.389.1419L} Li Y., Mo H.~J., Gao L., 2008,
MNRAS, 389, 1419

\bibitem[\protect\citeauthoryear{Lin, Jing,
\& Lin}{2003}]{2003MNRAS.344.1327L} Lin W.~P., Jing Y.~P., Lin L.,
2003, MNRAS, 344, 1327

\bibitem[\protect\citeauthoryear{Ludlow et al.}{2008}]{2008arXiv0801.1127L}
Ludlow A.~D., Navarro J.~F., Springel V., Jenkins A., Frenk C.~S.,
Helmi A., 2008, arXiv, arXiv:0801.1127

\bibitem[\protect\citeauthoryear{Maulbetsch et
al.}{2007}]{2007ApJ...654...53M} Maulbetsch C., Avila-Reese V.,
Col{\'{\i}}n P., Gottl{\"o}ber S., Khalatyan A., Steinmetz M.,
2007, ApJ, 654, 53

\bibitem[\protect\citeauthoryear{Mo, Jing, \&
B\"orner}{1997}]{1997MNRAS.286..979M} Mo H.~J., Jing Y.~P., Borner
G., 1997, MNRAS, 286, 979


\bibitem[\protect\citeauthoryear{Mo, Jing \& White}{1997}]
{1997MNRAS.284..189M}
Mo H.~J., Jing Y.~P., White S.~D.~M., 1997, MNRAS, 284, 189

\bibitem[\protect\citeauthoryear{Mo \& White}{1996}]{1996MNRAS.282..347M}
Mo H.~J., White S.~D.~M., 1996, MNRAS, 282, 347

\bibitem[\protect\citeauthoryear{Peacock \&
Smith}{2000}]{2000MNRAS.318.1144P} Peacock J.~A., Smith R.~E.,
2000, MNRAS, 318, 1144

\bibitem[\protect\citeauthoryear{Sandvik et
al.}{2007}]{2007MNRAS.377..234S} Sandvik H.~B., M{\"o}ller O., Lee
J., White S.~D.~M., 2007, MNRAS, 377, 234

\bibitem[\protect\citeauthoryear{Sheth, Mo,
\& Tormen}{2001}]{2001MNRAS.323....1S} Sheth R.~K., Mo H.~J.,
Tormen G., 2001, MNRAS, 323, 1

\bibitem[\protect\citeauthoryear{Sheth
\& Tormen}{1999}]{1999MNRAS.308..119S} Sheth R.~K., Tormen G.,
1999, MNRAS, 308, 119

\bibitem[\protect\citeauthoryear{Seljak \&
Warren}{2004}]{2004MNRAS.355..129S} Seljak U., Warren M.~S., 2004,
MNRAS, 355, 129

\bibitem[\protect\citeauthoryear{Wang et al.}{2005}]{2005MNRAS.364..424W}
Wang H.~Y., Jing Y.~P., Mao S., Kang X., 2005, MNRAS, 364, 424

\bibitem[\protect\citeauthoryear{Wang, Mo,
\& Jing}{2007}]{2007MNRAS.375..633W} Wang H.~Y., Mo H.~J., Jing
Y.~P., 2007, MNRAS, 375, 633

\bibitem[\protect\citeauthoryear{Wang et al.}{2008}]{2008arXiv0812.3723W}
Wang Y., Yang X., Mo H.~J., van den Bosch F.~C., katz N., Pasquali
A., McIntosh D.~H., Weinmann S.~M., 2008, arXiv, arXiv:0812.3723

\bibitem[\protect\citeauthoryear{Wechsler et
al.}{2006}]{2006ApJ...652...71W} Wechsler R.~H., Zentner A.~R.,
Bullock J.~S., Kravtsov A.~V., Allgood B., 2006, ApJ, 652, 71

\bibitem[\protect\citeauthoryear{Wetzel et al.}{2007}]{2007ApJ...656..139W}
Wetzel A.~R., Cohn J.~D., White M., Holz D.~E., Warren M.~S.,
2007, ApJ, 656, 139

\bibitem[\protect\citeauthoryear{Yang, Mo, \& van den
Bosch}{2003}]{2003MNRAS.339.1057Y} Yang X., Mo H.~J., van den
Bosch F.~C., 2003, MNRAS, 339, 1057

\bibitem[\protect\citeauthoryear{Zhu et al.}{2006}]{2006ApJ...639L...5Z}
Zhu G., Zheng Z., Lin W.~P., Jing Y.~P., Kang X., Gao L., 2006,
ApJ, 639, L5

\end{thebibliography}
\end{document}